\documentclass[twocolumn,english,prl,showpacs,floatfix]{revtex4}
\usepackage[T1]{fontenc}
\usepackage[latin1]{inputenc}
\usepackage{graphicx}
\usepackage{amssymb}
\newcommand \bea{\begin{eqnarray}}
\newcommand \eea{\end{eqnarray}}

\newcommand \la{\raisebox{-.5ex}{$\stackrel{<}{\sim}$}}
\newcommand{\av}[1]{\langle{#1}\rangle}

\makeatletter

\usepackage{graphicx}

\usepackage{babel}
\makeatother
\begin{document}

\title{Super-shell structures and pairing in ultracold trapped Fermi gases}

\author{Magnus \"{O}gren$^{1}$ and Henning Heiselberg$^{2}$}

\affiliation{$^{1}$Mathematical Physics, Lund Institute of Technology, P.O. 
Box
118, SE-22100 Lund, Sweden \\
$^{2}$University of Southern Denmark, Campusvej 55, DK-5230 Odense
M, Denmark}

%\date{\today{}}
\date{April 3, 2007}

\begin{abstract}
We calculate level densities and pairing gaps for an ultracold dilute
gas of fermionic atoms in harmonic traps under the influence
of mean field and anharmonic quartic trap potentials. Super-shell
structures, which were found in Hartree-Fock calculations, are calculated analytically within periodic orbit theory as well as from
WKB calculations.  For attractive interactions, the underlying level
densities are crucial for pairing and super-shell structures in gaps
are predicted.

\end{abstract}


\maketitle

Ultracold atomic gases have recently been used to create novel quantum
many-body systems such as strongly interacting high temperature
superfluids of fermions, Bose-Einstein condensates, Mott insulators in
optical lattices, etc.  These lab phenomena have a strong overlap with
condensed matter \cite{BCS}, nuclear \cite{BM} and neutron star
physics \cite{Pines}.  Finite fermion systems such as atoms in traps,
nuclei, helium and metal clusters, semiconductor quantum dots,
superconducting grains, etc., have additional interesting quantum
structures such as level spectra, densities and pairing.  These will
be observable as temperatures are further lowered in atomic trap
experiments. The high degree of control over physical parameters,
including interaction strength and density, makes the atomic traps
marvelous model systems for general quantum phenomena.

The purpose here is to calculate the level spectra, densities and
pairing for zero-temperature Fermi gases in harmonic oscillator (HO)
traps with anharmonic and mean field perturbations, and to show that
novel {\it super-shell} structures appear in both level densities and
pairing. In calculating level spectra by analytical periodic orbit
theory and WKB as well as numerical Hartree-Fock, we
also relate these different theoretical approaches to one another.

We treat a gas of $N$ fermionic atoms of mass $m$ in a
HO potential at zero temperature, interacting
via a two-body interaction with s-wave scattering length $a$.
We shall mainly discuss a spherically symmetric trap and a dilute gas
(i.e. where the density $\rho$ obeys the condition
$\rho|a|^3\ll 1$) of particles
with two spin states of equal population. The Hamiltonian is then given by
\bea  \label{H}
  H &=& \sum_{i=1}^{N} \left( \frac{{\bf p}_i^2}{2m} +
    \frac{1}{2} m\omega^2 {\bf r}_i^2  + U({\bf r}_i) \right) \,,
\eea
We will consider both external anharmonic potentials of the form $U=\varepsilon 
r^4$
and particle interactions:
$U({\bf r}_i)=(2\pi\hbar^2a/m)\sum_{j\ne i}\delta^3({\bf r}_i-{\bf r}_j)$.
When interactions are weak, the latter can be approximated by the mean field 
potential
\bea\label{UMF}
U(r) =\, \frac{2\pi\hbar^2a}{m} \rho(r) \,.
\eea
For a large number of particles and $U=0$ the Fermi energy is
$E_F=\tilde{n}_F\hbar\omega$ where $n_F=\tilde{n}_F-3/2 \simeq(3N)^{1/3} $ is the HO quantum 
number at the Fermi surface.
The HO shells are highly degenerate with states having angular momenta
$l=n_F,n_F-2,...,mod(n_F,2)$, due to the
$U(3)$ symmetry of the 3D spherically symmetric HO potential.
However, interactions split this degeneracy.
In the Thomas-Fermi (TF) approximation (see, e.g., \cite{PS})
the Fermi energy is
\bea
  E_F=\frac{\hbar^2k_F^2(r)}{2m}+\frac{1}{2}m\omega^2r^2+U(r) \, .
\eea
The density $\rho(r)=k_F^3(r)/3\pi^2$ is
\bea\label{rho}
    \rho(r) &=& \rho_0 \left(1-r^2/R_{TF}^2\right)^{3/2} \,,
\eea
inside the cloud $r\le R_{TF}=a_{osc}\sqrt{2\tilde{n}_F}$, where
$\rho_0=(2\tilde{n}_F)^{3/2}/3\pi^2a_{osc}^3$ is the central density
\cite{TF3}. For convenience we set
the oscillator length $a_{osc}=\sqrt{\hbar/m\omega}=1$ in the following.

Taylor expanding the density and thereby also the mean field of Eq. (\ref{UMF}) 
around the center gives
\bea\label{rhoT}
    \rho(r) &\simeq&
\rho_0 \left(1-\frac{3}{2}r^2/R_{TF}^2+\frac{3}{8}r^4/R_{TF}^4+...\right) 
\,,
\eea
the first term will
simply incorporate a constant shift in energies whereas the term
quadratic in radius renormalizes the HO frequency as
$\omega_{{\it eff}}=\omega\sqrt{1-6\pi a\rho_0/R_{TF}^2}$.
The third term is quartic in radius and is therefore also of the same form as the 
external potential
\bea\label{U4}
  U(r)\simeq \varepsilon r^4 \,,
\eea
with $\varepsilon=(3\pi\hbar^2a/4m)\rho_0/R_{TF}^4$.
Both the pure quartic potential and the mean field potential of 
Eq.~(\ref{UMF})
are anharmonic and change the level density by splitting the $l$
degeneracy of the HO shell $n_F$ at the Fermi surface.

We will now calculate analytically the level spectra from perturbative 
periodic orbit theory
for the quartic potential
and subsequently within semiclassical WKB wavefunctions
for both the quartic and the mean field potential of 
Eq.~(\ref{UMF}).
We will start with repulsive interactions where pairing is not present.

In periodic orbit theory \cite{BB}, the level density can be written
(to leading order in $\hbar^{-1}$) in terms of a perturbative HO trace formula 
\cite{Brack,Creagh}
\bea\label{gpt}
g(E)=\frac{E^{2}}{\left(\hbar\omega\right)^{3}}
     \left( 1+ Re\sum_{k=-\infty}^{\infty}\left(-1\right)^{k}{\cal M}\,
    e^{i2\pi kE/\hbar\omega} \,\right).
\eea
For the unperturbed HO ($U=0$) the modulation factor is
${\cal M}=1$. For a quartic perturbed potential, as in Eq.~(\ref{U4}),
the modulation factor was calculated in  \cite{Brack}
\bea
{\cal M}= \frac{\hbar}{k\sigma}
\left(e^{-i2k\sigma/\hbar-i\pi/2}+e^{-i3k\sigma/\hbar+i\pi/2}\right) \,,
\eea
with $\sigma=\varepsilon\pi E^2/\hbar^2\omega^3$, being a small
classical action. The two terms arise from
the change in actions for the circle and diameter orbits respectively
due to the quartic potential \cite{Brack}.
The resulting level density can be written in the factorised
form \cite{Ogren}

\begin{displaymath}
g(E) = \frac{E^2}{(\hbar\omega)^3} + \frac{4}{\pi\varepsilon} 
\sum_{k=1}^\infty
   \frac{\left( -1 \right) ^k}{k} \times
\end{displaymath}

\vspace{-4mm}

\bea \label{g}
\cos\left(  \frac{k}{\hbar}  \left(  \frac{2\pi E}{\omega}-\frac{5}{2}\sigma  \right)
\right) 
     \sin\left( \frac{ k\sigma }{2\hbar} \right)  \,.
\eea

Here, the first term is the average level density,
the cosine factor gives the rapid HO shell oscillations (modified by 
the perturbation)
which, however, are slowly modulated by the sine factor resulting in a 
beating pattern.  Moreover, the 
non-perturbed HO limit, equivalent
to ${\cal M}=1$ in Eq.~(\ref{gpt}), is recovered in the limit of 
$|\varepsilon| \rightarrow 0$, where the $U\left(3\right)$ symmetry is
restored.
The $k=1$ term in Eq.~(\ref{g}) gives the major oscillations in the level 
density
and is shown in Fig.~1 (a). The beating pattern or {\it super-shells} is 
clearly
observed.
The shell oscillations vanish when the argument of the sine in 
Eq.~(\ref{g}) is an integer
$S=1,2,3,...$ times $\pi$, i.e. $|\varepsilon| E^2/2(\hbar\omega)^3=S$.
This gives the {\it supernode} condition
\bea \label{sn}
   n_F=E/\hbar\omega = \sqrt{2S\hbar\omega/|\varepsilon|} \,.
\eea

\begin{figure}
\includegraphics[scale=0.44]{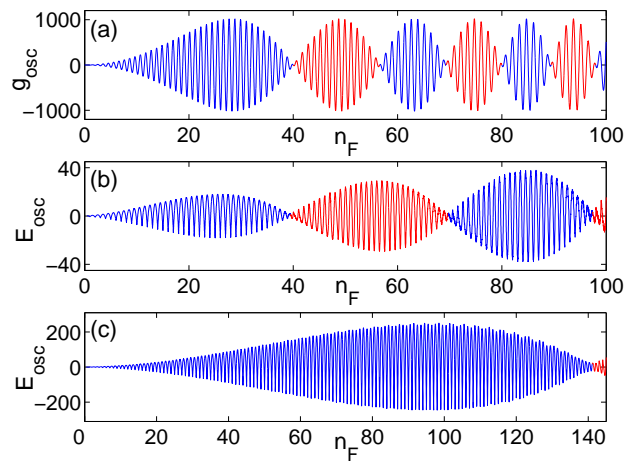}
\caption{(color online)
The upper figure (a) shows the leading term ($k=1$)
of the oscillating part of the perturbative level density of
Eq.~(\ref{g}) as a function of $n_F=E/\hbar\omega$, for the case of an
external potential $V=V_{HO}+\varepsilon r^4$ with
$\varepsilon=2/40^2\approx 0.0013$. The middle and lower figures (b,c) show the oscillating part of the
total energy according to a numerical HF calculation \cite{Yu}, with
interaction strength $2\pi a=1$ and $2\pi a=0.1$, as a function of the
HO shell number ($\hbar=\omega=1$). This illustrates qualitatively
that a supernode, e.g. at $n_F=40$, can be due to interaction (b) and/or an additional quartic term to the HO potential (a).}
\label{fig1}
\end{figure}

We now turn to an alternative calculation of the level density with WKB.
The splitting of the HO shells
degenerate levels $l=n_F,n_F-2,...,mod\left(n_F,2\right)$ in
the shell $n_F$ by the mean-field potential
can be calculated perturbatively in the dilute limit.
An excellent approximation for the radial HO wave function
with angular momentum $l$ and $(n_F-l)/2$ radial nodes in the HO shell
when $n_F\gg1$ is the WKB one \cite{HM,Heiselberg}:
\bea
   {\cal R}_{n_Fl}(r)\simeq \frac{2}{\sqrt{\pi}}
   \frac{\sin(k_l(r)r+\theta)}{k_l^{1/2}(r)r} \,, \label{RWKB}
\eea
between turning points $r^2_\pm=\tilde{n}_F\pm\sqrt{\tilde{n}_F^2-l(l+1)}$.
Here, $\tilde{n}_F=n_F+3/2$ and the WKB wave number $k_l(r)$ is
\bea \label{kl}
k_l^2(r)=2\tilde{n}_F-r^2-l(l+1)/r^2 \,.
\eea
When $n_F\gg 1$ the wave function has many nodes $1\ll l\ll n_F$ and
the oscillations in ${\cal R}_{nl}^2(r)$ can be averaged 
$\av{\sin^2(k_l(r)r)}=1/2$
\cite{HM}. The phase $\theta$ is then unimportant.
The single-particle energies for the anharmonic potential of
Eq.~(\ref{U4}) are simply
\bea \label{E}
  E_{n_F,l}&-&\tilde{n}_F\hbar\omega \,=
  \int_{r_-}^{r_+} U(r) |{\cal R}_{n_Fl}(r)|^2 r^2dr  \\
&=& \varepsilon\int_{r_-}^{r_+} \frac{2 r^4}{\pi k_l(r)} dr
  \,=\, \frac{\varepsilon}{2} \left[3\tilde{n}_F^2-l(l+1)\right] . 
\label{Ee}
\eea
It is special for the quartic perturbation that the level energies
are linear in $l(l+1)$. The resulting level spacing increases as
$(2l+1)$ just as the level degeneracy for $SO(3)$ symmetry. Therefore the 
level density is
constant within the bandwidth
\bea \label{supernode}
   D \equiv |E_{n_F,l\equiv 0}-E_{n_F,l=n_F}|
   =\varepsilon n_F^2/2
\eea
on energy scales larger than $2D/n_F$ but smaller than $D$.
The level density vanishes between the bandwidths of two neighbouring $n$ 
shells and therefore
it generally has a strong oscillatory behavior as shown in 
Fig.~\ref{fig1}~(a).
Its amplitude is largest when $D\sim \hbar\omega/2$. However, when
$D\simeq \hbar\omega$ the level density is constant and the oscillatory
behavior vanishes. This phenomenon repeats when $D=S\hbar\omega$ since
the level spectra then overlap $S$ times.
With the bandwidth of Eq.~(\ref{supernode}) under this condition, we obtain
exactly the same supernode condition  as for periodic orbit theory,
Eq.~(\ref{sn}).
We conclude that Craig's perturbative periodic orbit theory \cite{Creagh} is 
in exact agreement with perturbative WKB for a quartically perturbed spherical 
symmetric HO in three dimensions.

We now turn to the slightly more complicated mean field potential
of Eq.~(\ref{UMF}). Its level spectrum can also be calculated from
the WKB wave functions of Eq.~(\ref{RWKB}). Inserting them in Eq.~(\ref{E}), 
we obtain
\bea \label{ea}
  E_{n_F,l} -\tilde{n}_F\hbar\omega &=&
  2/\left( 3\pi^2\right) a  \tilde{n}_F^{3/2} \hbar\omega\, I \,.
\eea
Here, the integral $I$ is
\bea \label{I}
I= \int^1_{-1} \left(1-x\sqrt{1-l(l+1)/\tilde{n}_F^2}\right)^{3/2}
     \frac{dx}{\sqrt{1-x^2}}
\eea
where $x=(r^2-\tilde{n}_F)/\sqrt{\tilde{n}_F^2-l(l+1)}$. This integral is 
$I=\pi$ for $l\simeq n_F$ and $I=8\sqrt{2}/3$ for  $l=0$.
The bandwidth is therefore
\bea\label{D}
  D= 2/\left( 3\pi^2\right) a  n_F^{3/2} 
\hbar\omega\,\left(8\sqrt{2}/3-\pi\right)\,.
\eea
Inserting this bandwidth
in the supernode condition  $D=S\hbar\omega$ gives
\begin{equation}\label{S}
\left(8\sqrt{2}/3-\pi\right)
 2/\left( 3\pi^2\right) a  n_{F}^{3/2}=S \,.
\end{equation}
For example in the case  $2\pi a=1$ the supernodes $S=1,2,3,..$ should
occur when $n_F\sim 28,44,58$, etc. The Hartree-Fock (HF)
calculations of the oscillating part of the total energy, which is 
proportional
to the level density at the Fermi level \cite{BB}, result in slightly higher 
supernodes, as in Fig 1 (b).
The differences arise because the WKB calculations are perturbative in
the interaction strength, whereas in the HF calculation the MF
potential $U$ includes a large scattering length which, e.g., leads to
corrections for the effective oscillator frequency.
Also for the purely quartic term the perturbative approach
underestimates the exact supernodes (see Fig.~3 of~\cite{Brack}). For weaker interactions  $2\pi a=0.1$, the first supernode $S=1$ should
occur at $n_F=130$ according to the condition of Eq.~(\ref{S}),
in closer agreement with the HF result of Fig.~\ref{fig1}~(c).

For comparison, the Taylor expansion of the mean field potential leads to
the supernode condition of Eq.~(\ref{sn})
with $\varepsilon=(3\pi\hbar^2a/4m)\rho_0/R_{TF}^4$.
It differs from Eq.~(\ref{S}) by the prefactor, which is $\sim34$\% smaller. 
It is a better approximation to expand e.g. around
$r=R_{TF}/2\sqrt{2}=\sqrt{n_F}/2$, where the corresponding
prefactor is only $\sim8$\% smaller, such that the supernode in Fig~1~(c) is predicted to $n_F=137$.
Now expanding $I$ of Eq.~(\ref{I}) for small $l\ll n_F$, one finds
\bea \label{Il}
  I&=&(8/3)\sqrt{2} -l^2/\sqrt{2}n_F^2 \,,
\eea
resulting in the level spectrum \cite{HM}
\bea \label{Elow}
E_{n_F,l}-\tilde{n}\hbar\omega&=& \frac{\sqrt{2}}{3\pi^2} a n_F^{3/2} \hbar\omega\,
\left[\frac{16}{3}-\frac{l(l+1)}{n_F^2}\right]   \,.
\eea
This level density is constant at low $l$ as for the
potential in Eq.~(\ref{Ee}). However, near $l\sim n_F$ the density of
levels is slightly smaller as can be seen from the bandwidth corresponding to
Eq.~(\ref{Elow}), which is $\sim 12$\% larger, for a given $n_F$, than the 
bandwidth of
Eq.~(\ref{D}).
That the level density is not completely constant within the bandwidth
has the effect that a small periodicity remains even at the super-shell
condition $D=S\hbar\omega$. Therefore the shell oscillations do not
disappear completely at the supernodes, as can be seen in Fig.~\ref{fig1}~(b,c),
whereas for the purely quartic case (a) the oscillations disappear completely at the
supernodes.

Most atomic traps are not spherical but cigar shaped (prolate) with
$\omega_z\la\omega_\perp$.  The unperturbed HO energies
$E=n_z\hbar\omega_z+n_\perp\hbar\omega_\perp$ will generally
lead to a constant level density for energy scales larger than
$\hbar\omega_z$ but smaller than $\hbar\omega_\perp$. When the oscillator
frequency ratio $\omega_\perp/\omega_z$ is a rational number, level 
degeneracies
and larger oscillations will occur on the scale $\hbar\omega_z$. Interactions
will, however, smear this level density. In any case,
super-shell structure is not expected as in the spherical symmetric case.
In very oblate traps $\omega_z\gg\omega_\perp$ the
mean field potential is effectively two-dimensional and quadratic, i.e.
it does not split the HO shells \cite{HM,Zyl}. Thus we may expect strong
oscillations in the level density on the scale $\hbar\omega_\perp$, but
again no super-shell structure.

Attractive interactions lead to pairing by an amount that is exponentially 
sensitive to the
underlying level density near the Fermi surface \cite{HM,BM,BH,Heiselberg}. 
The
level density is the same for repulsive and attractive interactions except 
that the levels
are reversed when the sign of $\varepsilon$ (Eqs.~(\ref{g})) and (\ref{E})) or $a$ is changed 
(Eq.~(\ref{ea})).
Therefore we can use the level densities and bandwidths calculated above for 
pairing calculations.
Pairing in finite systems is described by the Bogoliubov-de Gennes (BdG) 
equations
\cite{deGennes} and take place between time-reversed states.
As shown in
\cite{BH} these states can be approximated by HO wave functions in dilute
HO traps as long as the gap does not exceed the oscillator energy,
$\Delta\la\hbar\omega$. Solving BdG for such finite systems
is numerically complicated and we shall therefore apply further simplifying 
approximations, namely that the pairing gap $\Delta_{nl}$ and the wavefunction overlap 
matrix elements
vary slowly with level $l$
in a shell $n$. Both approximations
are fair for the trapped atoms as argued in \cite{Heiselberg} and deviations 
can be understood.
As result we arrive at a much simplified gap equation
\bea \label{Dgap}
1=\frac{G}{n_F^2} \int_0^{\sim 2n_F} \frac{g(E)\, 
dE}{\sqrt{(E-\mu)^2+\Delta(\mu)^2}} \,.
\eea
Here, the supergap
$G=32\sqrt{2n_F}|a|\hbar\omega/15\pi^2$ was calculated
in \cite{HM} as the pairing gap
when all states in a shell can pair; this is the case for
a region of interaction strengths and particle number where the gap is large 
as compared to
the level splitting, yet small compared to the shell splitting $\hbar\omega$. $\Delta(\mu)=\Delta_{nl}$~is the gap at the Fermi surface. $g(E)=n_F^2/D$~is the level density within each bandgap $D$ around every 
shell $n=0,1,...,\sim$$2n_F$
but vanishes between the bandgaps.
The gap equation thus reduces to
$1=(G/D)\sum_{n=0}^{\sim 2n_F} \int_0^D 
dE/\sqrt{(E + n\hbar\omega-\mu)^2+\Delta^2}$.
The chemical potential $\mu$ can be determined from the level spectrum; as 
we gradually fill
particles into the shell $n_F$ at the Fermi surface, $\mu$ 
increases from
$n_F\hbar\omega$ to $n_F\hbar\omega+D$.
The cut-off $n\la 2n_F$ in the sum of the gap equation models as a
first approximation the more rigorous regularization procedure
described in Ref.\ \cite{BruunBCS} that is required for a
delta-function pseudo-potential.

\begin{figure}
\includegraphics[scale=0.44]{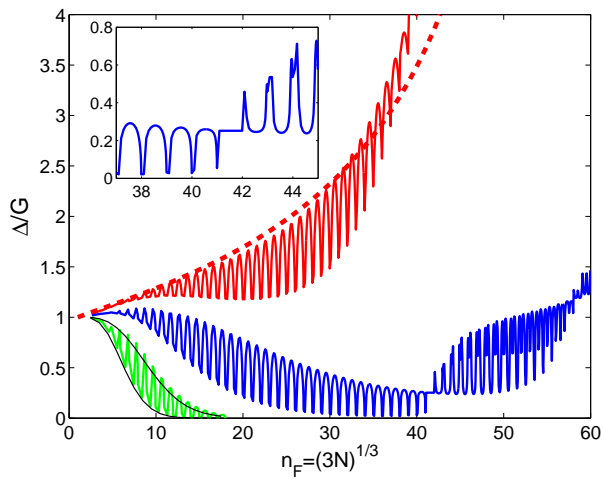}

\caption{(color online) Multi-shell pairing gaps for a HO trap with
an additional quartic term in the potential with $\varepsilon=2/40^2$,
i.e. for the level density of Fig.~1~(a) with supernodes at $n_F\simeq 40$, 
$40\sqrt{2}\approx57, etc.$
The interaction strength is $a=-0.05$ (top red curve),
$a=-0.03$ (middle blue curve, with the inset figure around the first supernode) and
$a=-0.01$ (lower green curve).
In the inset plot it is clearly seen that
the local minima for $l\sim n_F$ and $l\sim 0$ before the supernode
turns into local maxima after the supernode, as a consequence of
overlapping shells.
The dashed (red) line is the multi-shell gap 
$\Delta=G/(1-2G\ln(n_F)/\hbar\omega)$
for $a=-0.05$ and the upper/lower thin solid line (black) are
the single mid-/end-shell pairing for $a=-0.01$ (see text).}
\label{fig2}
\end{figure}

By solving this simplified gap equation of Eq.~(\ref{Dgap}),
we find that it still
contains and displays the essential interplay between the variation in
level density and pairing.  To illustrate the super-shell structure in
pairing, we take the strongly anharmonic trap potential used for the
level spectra in Fig.~1~(a), and calculate the pairing arising from a
weak attractive scattering length $a<0$.  For sufficiently weak
interactions such that pairing only takes place in the shell at the
Fermi surface, we obtain the expected result from the gap equation:
$\Delta=G$ when $D\ll\Delta$, whereas for $D\gg\Delta$ we get
$\Delta=D\exp(-D/2G)$ midshell ($\mu=n_F\hbar\omega+D/2$) and
$\Delta=2D\exp(-D/G)$ endshell ($\mu=n_F\hbar\omega$ or
$\mu=n_F\hbar\omega+D$). Pairing is thus stronger at midshell than
at endshell, where there are fewer states to pair \cite{Heiselberg}, and
strong shell oscillations follow as shown in Fig.~2. For stronger
interactions, pairing also takes place between states in shells around
the Fermi shell and Eq.~(\ref{Dgap}) gives:
$\Delta=G/\left(1-2G\ln\left(n_F\right)/\hbar\omega\right)$ for small
bandwidth \cite{BH}. In Fig.~2 this curve is compared with the
finite bandwidth result, which has strong oscillations except at the
supernodes where the level density is continuous. At a supernode
$D=\hbar\omega$ and the gap equation~(\ref{Dgap}) leads to a gap
$\Delta=2n_F\hbar\omega\exp(-\hbar\omega/2G)$ \cite{Heiselberg}.

In summary, level densities, shell-oscillations and super-shell structures in
anharmonic traps calculated from numerical Hartree-Fock and analytical
periodic orbit theory as well as WKB were found to match to leading
order. Analogous super-shell structures were found in pairing from an
approximated BdG calculation. The mean field in atomic nuclei also
have a large anharmonic potential and the HO shells start to overlap
(the first supernode) already for heavy nuclei with $n_F\sim 5-6$. The
interplay of level spectra and multishell pairing is, however,
difficult to disentangle in nuclear pairing due to strong spin-orbit
effect and small particle number. Ultracold atomic traps, however,
provide ideal systems for observing the rich quantum structures such as
level densities and pairing.

Discussions with Matthias Brack on periodic orbit theory, Ben
Mottelson on (nuclear) shell theory and pairing, and proof reading by Joel
Corney, are gratefully
acknowledged.


\begin{thebibliography}{10}
\bibitem{TF3} For a finite number of particles the
factor $\tilde{n}=n_F+3/2$ includes a correction to $n_F$, which
has been checked numerically to improve the TF approximation and
slightly change the prediction of supernodes.
\end{thebibliography}
\end{document}